# Reconstruction of hit-time and hit-position of annihilation quanta in the J-PET detector using the Mahalanobis distance

N. G. Sharma[1*], M. Silarski[1], T. Bednarski[1], P. Białas[1], E. Czerwiński[1], A. Gajos[1], M. Gorgol[3], B. Jasińska[3], D. Kamińsk[1], L. Kapłoń[1,2], G. Korcyl[1], P. Kowalski[4], T. Kozik[1], W. Krzemień[5], E. Kubicz[1], Sz. Niedźwiecki[1], M. Pałka[1], L. Raczyński[4], Z. Rudy[1], O. Rundel[1], A. Słomski[1], A. Strzelecki[1], A. Wieczorek[1,2], W. Wiślicki[4], M. Zieliński[1], B. Zgardzińska[3], P. Moskal[1]

[1] *Faculty of Physics, Astronomy and Applied Computer Science, Jagiellonian University,*
*S.Łojasiewicza 11, 30-0348, Kraków, Poland*
[2] *Institute of Metallurgy and Materials Science of Polish Academy of Sciences,*
*W. Reymonta 25, 30-059, Kraków, Poland*
[3] Department of Nuclear Methods, *Institute of Physics,Maria Curie-Sklodowska University,*
*Pl. M. Curie-Sklodowskiej 1, 20-031, Lublin, Poland*
[4] *Świerk Computing Centre, National Centre for Nuclear Research,*
*A. Soltana 7, 05-400 Otwock-Świerk, Poland*
[5] *High Energy Physics Division, National Centre for Nuclear Research,*
*A. Soltana 7, 05-400 Otwock-Świerk, Poland*

* email: pnp.neha@gmail.com

**Abstract**

The J-PET detector being developed at Jagiellonian University, is a Positron Emission Tomograph composed of the long strips of polymer scintillators. At the same time it is a detector system which will be used for studies of the decays of positronium atoms. The shape of photomultiplier signals depends on the hit-time and hit-position of the gamma quantum. In order to take advantage of this fact a dedicated sampling front-end electronics which enables to sample signals in voltage domain with the time precision of about 20 ps and novel reconstruction method based on the comparison of examined signal with the model signals stored in the library has been developed. As a measure of the similarity we use the Mahalanobis distance. The achievable position and time-resolution depends on number and values of the threshold levels at which the signal is sampled. A reconstruction method, as well as preliminary results are presented and discussed.

**Introduction**

A new concept of large acceptance Jagiellonian PET (J-PET) based on the polymer scintillators [1–9] has been developed recently, which provides a solution for a whole-body PET imaging with an improved Time of Flight (TOF) resolution[1] as it relies more on the precise time measurement than on energy resolution (like in traditional PETs) and the signals that we probe are very narrow few nanaoseconds. To this end a dedicated new electronics was developed which enables signal sampling in voltage domain with a time precision of about 20 ps [13]. Moreover, we have been developing new methods of gamma quanta hit time and position reconstruction to fully exploit potential of the new tomograph [11,14,15]. One of the developed methods is based on comparison of measured signals with respect to a library of synchronized model signals registered for a set of well-defined positions of scintillation points. The hit position is reconstructed as the one corresponding to the signal from the library which is most similar to the measured signal [11]. The measured signal can be compared also to the averaged model signals determined for each scintillation point which speeds up significantly the reconstruction [12]. The degree of similarity between the measured and the reference signals from the library can be expressed by several metrics e.g. the methods described in the articles [11,12] use $\chi^2$ metric. In the present article we discuss a method based on the Mahalanobis distance [16], which accounts for the correlation among the compared variables.

---

[1] In current tomographs the best TOF resolution was achieved with LSO crystals and it is equal to about 400 ps [10] and in our previous studies with the plastic scintillators ~290 ps was achieved [11, 12].

**Principle of time and position reconstruction using library of model signals**

The method was validated by using the experimental data collected by means of the double strip J-PET prototype and $^{22}$Na isotope as a source of annihilation gamma quanta. For noise suppression and selection of annihilation gamma quanta a coincident registration of signals from both detectors was used. Pedestal correction was also implemented to all the collected signals. In order to construct the library only with the high energy deposition events, only those signals were selected for which number of registered photoelectrons is greater than half of the number of photoelectrons corresponding to the Compton edge for 511 keV gamma quanta. Next, a two dimensional plot of number of photoelectrons obtained from left and right sided photomultipliers connected to the same strip were bisected into four regions (region 1, region 2, region 3 and region 4). An example for central position is shown in Fig. 1. Bisection was done in order to suppress the uncertainties arising due to the variation of time resolution as a function of number of registered photoelectrons. Then, for each bisected region signals were synchronized by shifting their time scales in such a way that time of the gamma quantum hit inside the detector is the same for all events in the library. Finally, the shape of a model signal for each bisected region is determined by averaging pedestal corrected and synchronized signals[2].

For reconstructing the hit position of gamma quanta in the detector the measured signals are compared with each model signal from the library. Signals can be represented as a vector of time values at defined threshold levels. In this article we consider two threshold case (-60 mV and -120 mV) where, $t_{L1}$, $t_{L2}$ and $t_{R1}$, $t_{R2}$ are time values on left and right signals (registered simultaneously at both ends of the scintillator) defined at -60 mV and -120 mV threshold levels, respectively:

---

[2] For detailed descriptions of experimental setup and the method of constructing library of model signals see ref. [12].

$$\vec{t} = [t_{L1}, t_{L2}, t_{R1} t_{R2}] \qquad (1)$$

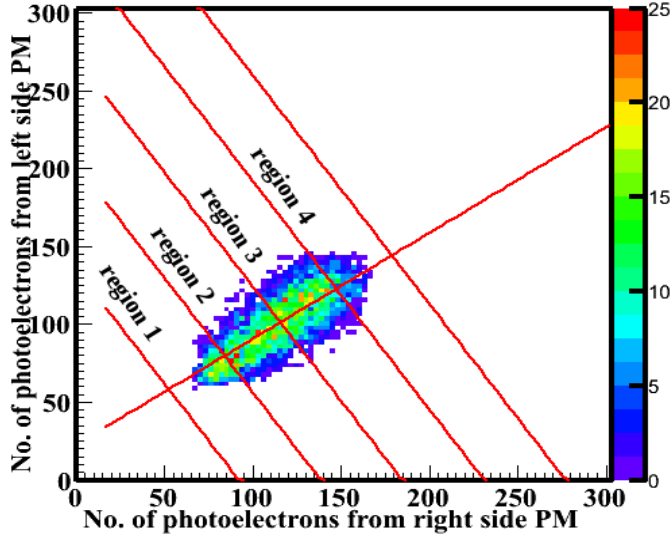

Fig 1: Distribution of number of photoelectrons obtained from left and right sided photo-multipliers connected to scintillator at central position.

The position and time of the examined signal is determined by minimizing the Mahalanobis distance (Eq. 2) between this signal and the model signals stored in the library. For each

$$M(z, \delta t, region) = \sqrt{[\vec{X}(\delta t, z, region)][Cov(z, region)]^{-1}[\vec{X}(\delta t, z, region)]} \qquad (2)$$

$$Cov_{(i,j)}(z, region) = \sum_{k=1}^{N} \frac{\left(\vec{t}_{(k(i))}(z, region) - \vec{t}_{avg(i)}(z, region)\right)\left(\vec{t}_{(k(j))}(z, region) - \vec{t}_{avg(j)}(z, region)\right)}{N} \qquad (3)$$

measured position a covariance matrix has been constructed using Eq. 3, where, k is an index that shows a current signal from the set of measured signals for given position, N is total number of measured signals. $\vec{t}_{k(i)}$ indicates an i$^{th}$ component of the k$^{th}$ signal and $\vec{t}_{avg(i)}$ denotes i$^{th}$ component of the average signal. $\vec{X}(\delta t, z, region)$ given in Eq. 4 is the difference between the vectors describing the examined and model signals with an additional term δt which is a parameter varied when minimizing the Mahalanobis distance:

$$\vec{X}(\delta t, z, region)_{(i)} = \vec{t}_{model(i)}(z, region) - \vec{t}_{k(i)} - \delta t \qquad (4)$$

The reconstructed hit position is the position of most similar signal from the library with respect to measured signal ( i.e. model signal for which Mahalanobis distance computed for different bisected region is minimal[3]). The time of particle interaction is determined as a relative time between the measured signal and the most similar model signal from the library. This method provides also determination of the gamma quantum time of flight (TOF) given by Eq 5:/

$$\begin{aligned} t_{firstStrip} &= \delta t_{firstStrip} \\ t_{secondStrip} &= \delta t_{secondStrip} \\ TOF &= t_{secondStrip} - t_{firstStrip} \end{aligned} \qquad (5)$$

---

[3] The degree of similarity is represented by the Mahalanobis distance value.

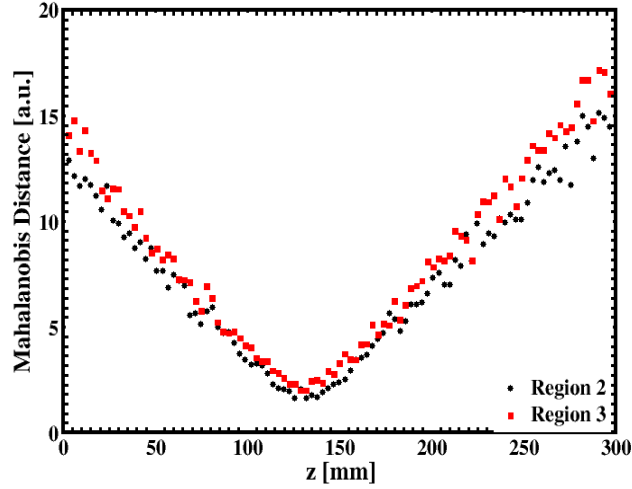

Fig 2: Distribution of Mahalanobis distance (defined in Eq. 3) for a signal from the central position (i.e. 150 mm) from region 2 and 3.

where $\delta t_{firstStrip}$ and $\delta t_{secondStrip}$ denotes shift in time obtained for two scintillators, for which the computed Mahalanobis distance defined in Eq. 2 is lowest.

**Preliminary results on performance of the method**

An example of the Mahalanobis distance distribution calculated according to Eq. 2 for one of the measured exemplary signal at central position of the strip is shown in Fig. 2. One can see a clear minimum corresponding to z ≈ 150 mm. Fig. 3a and b show distributions of differences between the true and reconstructed values for position and TOF, respectively[4]. The estimated resolution of position and TOF reconstruction based on these distributions, amount to $\sigma_z \approx 10$ mm and $\sigma_{TOF} \approx 140$ ps, respectively. They were determined for signals measured at several positions along the scintillator as it is shown in Fig. 4a and b. These results indicate that the resolutions do not change much with position.

---

[4] In principle the true value of TOF should be equal to zero when the source was positioned in the middle between detection modules. However, due to different electronics offsets the reconstructed mean values of TOF may be different from zero.

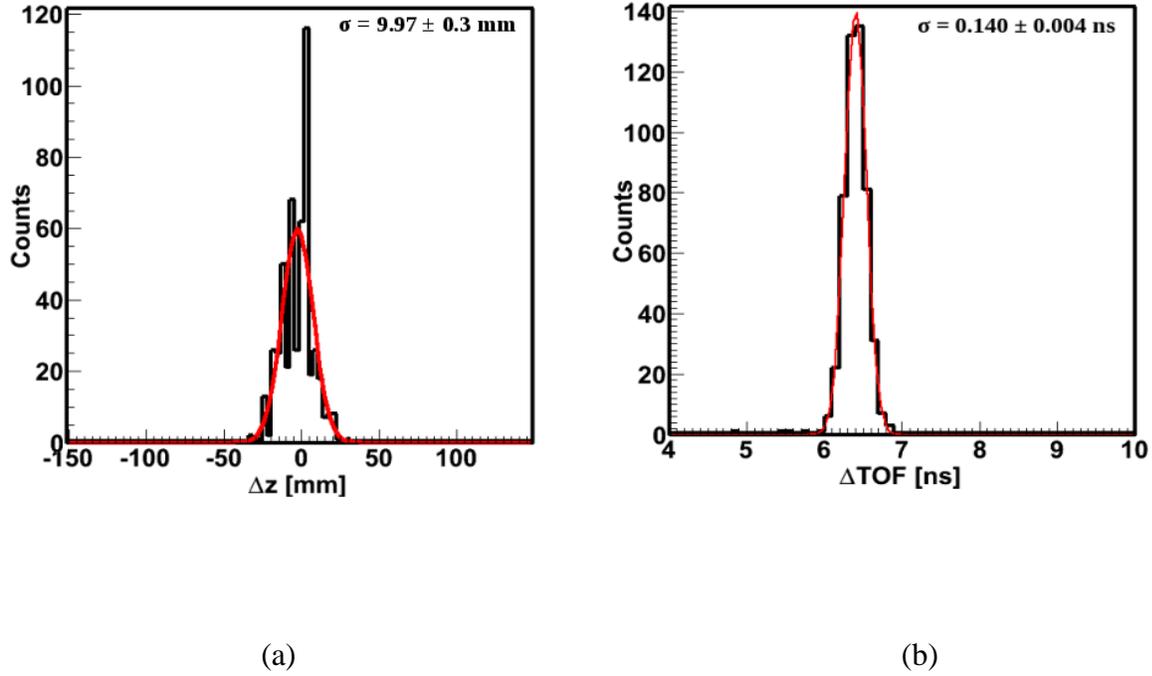

(a)          (b)

Fig 3: (a) Distribution of differences between the true and reconstructed position Δz for signals measured at z = 150 mm. (b) Distribution of differences between the true and reconstructed TOF for events registered at z = 150 mm.

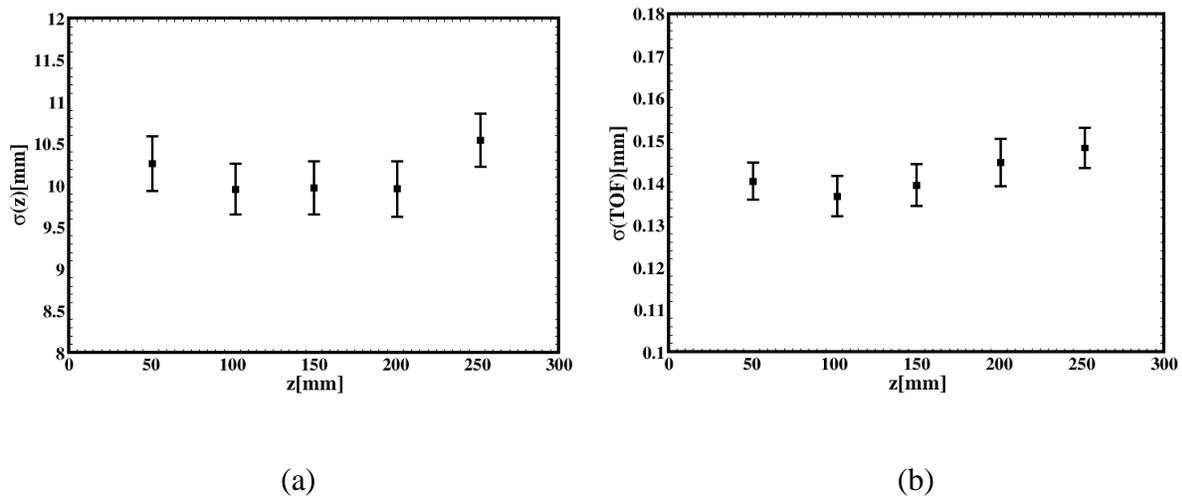

(a)          (b)

Fig 4: (a) Position resolution as a function of the position of gamma quantum interaction. (b) TOF resolution as a function of position along the scintillator.

**Summary**

The reconstruction method introduced in this article was validated for two-threshold levels and the preliminary results show that it is possible to obtain a spatial resolution of about 10 mm (σ) for the gamma quanta hit position, and TOF resolution of about 140 ps (σ). In the present version of the J-PET which is being built now we install electronics which will allow us to determine time at four-threshold levels, so a further improvement is expected in the future by including more bisected regions and more threshold levels.